# The Author Is Sovereign: A Manifesto for Ethical Copyright in the Age of AI

Ricardo Fitas
ricardo.fitas@tu-darmstadt.de
rfitas99@gmail.com
ORCID: 0000-0001-5137-2451

In an age dominated by Artificial Intelligence (AI), algorithmic distribution, and mass content scraping, the notion of authorship is being steadily eroded (Edwards, 2023). Institutions and platforms are increasingly dismantling the ethical contract between creators and the public by rapidly harvesting copyrighted works to train machine learning models and by treating "fair use" as a convenient shortcut to bypass consent and access protected work.

Copyright ethics have become a battleground of clashing ideologies. Creative Commons promotes open licensing as a tool for cultural sharing, often privileging moral virtue over economic value. Richard Stallman, by contrast, advocates for software freedom while rejecting strong copyright protections. Ayn Rand, on the other end of the spectrum, defends intellectual property as a cornerstone of individual liberty. Yet despite these contrasting views, society continues to address the symptoms of a copyright system corroding from within, rather than confronting its root cause.

AI models are criticized only after they have scraped and repurposed content without consent. In response, institutions scramble to impose so-called "solutions" that may be even more damaging than the original violations: Open Access regimes that mandate Creative Commons licensing without offering transparent or equitable returns to authors; or punitive measures against fraudulent publishing practices that ignore the deeper incentive structures created by copyright erosion. These reactions attempt to superficially compensate for systemic failures without ever acknowledging that the crisis stems from the deliberate dilution of authorship rights. Instead of restoring control to creators, institutions continue to reinforce the commodification of creative labor.

Modern digital ecosystems are designed around ease of access and frictionless reproduction. While this may appear to democratize content consumption, doesn't it strip the creator of value – and of agency? At the core of this apparatus, what passes for "fair use", in its current legal interpretations, has become a gold-plated loophole that benefits institutions, platforms, and corporations at the expense of individual authors, rendering them increasingly irrelevant. Meanwhile, governments and publishers continue to frame Open Access as a moral imperative without providing adequate structural incentives. As a result, the author's role is eroded, and creators are absorbed into systems that profit from the very content they undercompensate or misattribute. In the context of AI training, this erosion becomes even more critical: data is scraped, repackaged, and redeployed without consent, under the assumption that cultural production belongs to the commons by default. It does not.

I actually believe in a radically different framework: **one in which the author is sovereign over their own work—including the right to govern its use.** Below, I outline seven principles that move us from a culture of exploitative appropriation toward one of ethical negotiation - placing the burden of justification on users, platforms, and governments, not on creators themselves:

**1. The author retains unconditional control over their work by default.** Ownership comes with creation. No institutional, governmental, or collective presumption may override it without explicit consent. This principle affirms that the act of creation itself establishes a moral relationship between the author and their work—one that cannot be broken or diluted by secondary actors without negotiation. We live in a digital world where creative outputs are increasingly modular, remixable, and algorithmically disseminated, which only heightens the need for authors to retain default control. It guards against silent appropriation under vague appeals to "public interest" or "technological advancement," ensuring that creators are not reduced to invisible inputs within extractive systems. Cultural progress demands that every act of reuse be treated as an opportunity for dialogue, consent, and compensation.

**2. "Fair use" shall not apply by default.** Any usage of a protected work—especially in digital environments—must be explicitly permitted by the author. Fair use was originally conceived as a narrow legal doctrine to enable criticism, education, or parody, but it has become a legal grey zone exploited at scale. Today, virtually anything can be claimed as "transformative," and the author's voice is displaced by loopholes and corporate interpretations. This shift represents not just a distortion of law, but an ethical betrayal. In its modern digital form, fair use is no longer an exception—it has become an unregulated assumption. If reuse is to be legitimate, it must be transparent, consent-based, and negotiated directly with the creator.

**3. If governments or institutions seek to implement "fair use"-like access to protected works, they must negotiate equitable compensation or subsidies with the original author.** The public good is not a license for exploitation. While access to knowledge and culture is a valid societal goal, it cannot be pursued through uncompensated appropriation. Any attempt by governments, universities, or AI developers to build systems of open access, data sharing, or algorithmic training using protected works must begin with a single premise: the author's right to be asked, acknowledged, and fairly compensated. Reasonable, isn't it?

Calls for universal accessibility are often framed as progressive, but they mask a deeper contradiction: access is declared a public right; creation, a private burden. It is ethically indefensible to fund technological infrastructure with public money while refusing to fund the creative inputs that power it. If society truly values openness, it must stop powering the engine while starving the maker. Because we cannot keep starving the maker.

Cultural redistribution must be contractual, not extractive—rooted in licensing, subsidies, or public procurement that recognizes authors as rightful stakeholders.

**4. Copyright ownership must result from explicit, transparent negotiation with the author.** Copyright is a declaration of authorship and creative agency, not a procedural formality. As such, ownership must never be assumed, institutionally inherited, or buried in opaque publishing workflows. Every single transfer of copyright must be the result of a conscious, informed negotiation with the author. It must be documented, voluntary, and accompanied by clear terms, including duration, scope, financial arrangements, and rights of reversion. This principle challenges prevailing norms in academia, publishing, and digital content platforms, where authors are told they must trade ownership for visibility, give up control for participation, and relinquish agency for relevance.

**5. Open Access licensing must be driven by tangible incentives—not moral appeals alone.** Authors may choose to license their work openly, but that choice must remain voluntary, informed, and meaningfully rewarded. An ethical Open Access ecosystem cannot depend on vague promises of increased readership. It must be grounded in reciprocity. This means offering authors tangible benefits: funding for publication, formal recognition in evaluations, visibility metrics that advance careers, and systems for revenue sharing or reuse royalties when work is used commercially.

Moreover, voluntary Open Access cannot coexist with policies that make it mandatory—whether through funder mandates, publisher coercion, or institutional pressure. Licensing one's work openly should be a conscious act of generosity — not a quiet surrender of worth.

**6. If the author retains copyright, any value derived from their work must be proportionally shared—not captured by governments, platforms, or publishers.** Authorship is not symbolic; it is a value-generating act. When a work creates visibility,

revenue, or institutional prestige—whether through licensing, advertising, AI training, or citations—the author has a legitimate claim to part of that value. Yet governments, publishers, and platforms often extract this value while excluding creators from the reward structure. This principle is simple: if others profit, the author must benefit. That benefit may take many forms—royalties, reuse fees, or metrics—but it must never be zero. To deny this is to break the basic contract of creativity. Authors must also retain the right to create, join, or exit collective arrangements on their own terms. Guilds, cooperatives, or professional societies may offer meaningful alternatives for negotiating fairer deals at scale—but participation must always remain voluntary, author-led, and reversible. Sovereignty means more than the right to say no—it includes the freedom to collaborate without coercion.

**7. The public domain must be ethically curated, not assumed.** The public domain is often romanticized as neutral cultural heritage, but it is a political construct shaped by power and precedent. No work belongs there by default, fiat, or under vague ideals of openness. Authorship is a lifelong bond — not a temporary one — between creator and creation. Entry into the public domain should occur only through the natural expiration of copyright or the author's informed, voluntary decision.

This challenges the assumption that accessibility implies public ownership, and resists moves to declare entire content categories—like AI training data or educational materials—as communal without consent. Again: a true commons must be built—not claimed—through consent, attribution, and respect.

The benefits of this shift would be profound. Stronger recognition of authorial sovereignty would dismantle ghostwriting, honorary authorship, and gift-name practices—the quiet distortions that erode the ethics of creation. With clear ownership norms, paper mills would lose their market, as only the true creator could claim rights and receive incentives. Copyright reform centered on ethical negotiation would also clarify boundaries, making it harder for bad actors to exploit ambiguities around reuse, citation, or data scraping.

Perhaps most importantly, by valuing the author directly, funding models would shift away from opaque lobbying structures and agency-driven metrics, toward systems that invest transparently in original work. Art would flourish in that shift—not by chance, but by design. Art, born of the creator's sovereign will, rooted in the very impulse that makes us human. It is this art—not derivative, not scraped, but authored— that fuels true innovation, and stands inimitable.

This manifesto does not seek to dictate what any creator must do with their work. Authorial sovereignty also includes the right to abdicate that sovereignty. If a creator wishes to waive their rights, adopt open licenses, or donate their creation to the public domain out of personal conviction, that choice must be equally respected—so long as it is informed, voluntary, and not imposed by institutional norms or collective moral pressure. True freedom includes the right to choose generosity, sharing, or even submission, so long as it is never turned into a rule for others.

Ethical copyright is not about restriction—it is about responsibility. A society that wishes to build with the works of others must be prepared to negotiate, compensate, and recognize. In the age of AI, the stakes are even higher. Creative sovereignty is not a privilege—it is the precondition for cultural legitimacy. To treat it as anything less is to betray the foundation of authorship itself.

The time has come to abandon the rhetoric of exception and embrace a future rooted in voluntary exchange, contractual clarity, and authorial dignity.

*I wrote this because I am still sovereign. I might not always be able to.*